\documentstyle[preprint,aps,epsf]{revtex}
\begin{document}
\draft
\title{Commensurability in One Dimension and the Josephson Junction Ladder}
\author{Colin Denniston$^{1,2,}$\cite{present} and Chao Tang$^{2}$}
\address{$^{1}$Department of Physics, Princeton University, 
Princeton, New Jersey 08544}
\address{$^{2}$NEC Research Institute, 4 Independence Way, Princeton,
New Jersey 08540}

\date{\today}
\maketitle

\begin{abstract}
We study a Josephson junction ladder in a magnetic field in the absence of 
charging effects via a transfer matrix formalism.  The eigenvalues of the 
transfer matrix are found numerically, giving a determination of the 
different phases of the ladder.  The spatial periodicity of the ground state 
exhibits a devil's staircase as a function of the magnetic flux filling factor 
$f$.  If the transverse Josephson coupling is varied a continuous 
superconducting-normal transition in the transverse direction is observed, 
analogous to the breakdown of the KAM trajectories in dynamical
systems.  We also examine how these properties may be affected by
a current injected along the ladder. 
\end{abstract}

\narrowtext
\section{Introduction}
\begin{figure}
\narrowtext
\centerline{\epsfxsize=3in
\epsffile{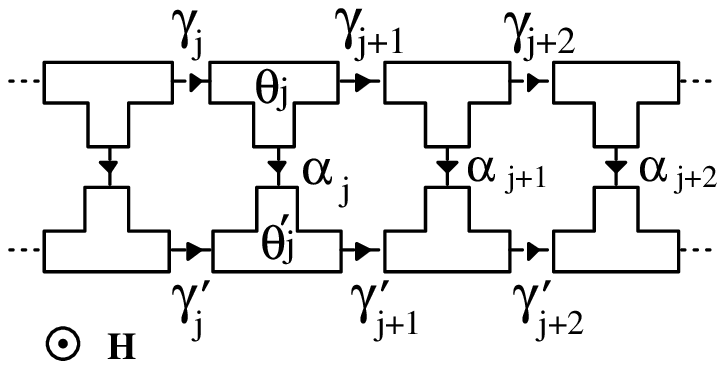}}
\vskip 0.1true cm
\caption{The Josephson junction ladder is formed by the 
arrangement of the super\-conducting islands.  The field ${\bf H}$ is out of 
the page and the arrows indicate the direction of the gauge invariant phase 
differences.}
\label{ladder}
\end{figure}

Two-dimensional arrays of Josephson junctions have attracted much recent 
theoretical and experimental attention \cite{jja}.  Interesting physics arises 
as a result of competing vortex-vortex and vortex-lattice interactions.  It is 
also considered to be a convenient experimental realization of the frustrated 
XY models.  In this paper, we expand and elaborate on our previous letter 
\cite{us1} on the simplest of such system, namely the Josephson junction ladder
 (JJL) \cite{kardar,laddergroups,mooij} shown in Figure~\ref{ladder}.   

To construct the system, superconducting elements are placed at the ladder 
sites.  Below the bulk super\-conducting-normal transition temperature, the 
state of each element is described by its charge and the phase of the 
superconducting wave function \cite{anderson}.  In this paper we neglect 
charging effects, which corresponds to the condition that $4e^2/C \ll
J$, with $C$ being the capacitance of the element and $J$ the Josephson
coupling.  Let $\theta_j$ ($\theta_j'$) denote the phase on the 
upper (lower) branch of the ladder at the $j$'th rung.  The Hamiltonian 
for the array \cite{tinkham} can be written in terms the gauge invariant 
phase differences, $\gamma_j=\theta_{j}-\theta_{j-1}-(2\pi/\phi_0) 
\int_{j-1}^j A_x dx$, $\gamma_j'=\theta_{j}'-\theta_{j-1}'-(2\pi/\phi_0)
\int_{j'-1}^{j'} A_x dx$, and $\alpha_j=\theta_{j}'-\theta_{j}
-(2\pi/\phi_0) \int_{j}^{j'} A_y dx$:
\begin{equation}
{\cal H} = - \sum_j (J_x\cos\gamma_j + J_x\cos\gamma_{j}' 
+ J_y\cos\alpha_j),
\label{ham1}
\end{equation}
where $A_x$ and $A_y$ are the components of the magnetic vector potential
along and transverse to the ladder, respectively, and $\phi_0$ the flux
quantum.  The sum of the phase 
differences around a plaquette is constrained by 
\begin{equation}
\gamma_j-\gamma_{j}'+ \alpha_{j}-\alpha_{j-1}=2 \pi (f-n_{j}),
\label{constraint}
\end{equation} 
where $n_j=0, \pm 1, \pm 2, \cdots$ is the vortex occupancy number and 
$f=\phi/\phi_0$ with $\phi$ being the magnetic flux through a plaquette.  With 
this constraint, it is convenient to write Eq.~(\ref{ham1}) in the form 
\begin{eqnarray}
{\cal H} &=&-J\sum_j \{2\cos\eta_j\cos[(\alpha_{j-1}-\alpha_{j})/2+
           \pi(f-n_j)]  \nonumber\\
         & & \quad\quad\quad \mbox{}+ J_t\cos\alpha_j\},
\label{ham2}
\end{eqnarray}
where $\eta_j=(\gamma_j+\gamma_j')/2$, $J=J_x$ and $J_t=J_y/J_x$.   The 
Hamiltonian is symmetric under $f\rightarrow f+1$ with $n_j\rightarrow n_j+1$,
 and $f\rightarrow -f$ with $n_j\rightarrow -n_j$, thus it is sufficient to 
study only the region $0 \le f \le 0.5$.  Since in one dimension ordered 
phases occur only at zero temperature, the main interest is in the ground 
states of the ladder and the low temperature excitations.  
Note that in Eq.~(\ref{ham2}) $\eta_j$ decouples from $\alpha_j$ and $n_j$, so 
that all the ground states have $\eta_j=0$ to minimize $\cal H$.
The ground states will be among the solutions to the current conservation 
equations $\partial {\cal H}/\partial \alpha_j=0$: 
\begin{eqnarray}
J_t\sin\alpha_j &=&
\sin[(\alpha_{j-1}-\alpha_{j})/2+\pi(f-n_j)] \nonumber\\
                 & & \mbox{} - \sin[(\alpha_j-\alpha_{j+1})/2+\pi(f-n_{j+1})].
\label{ccons}
\end{eqnarray}
For any given $f$ there are a host of solutions to Eq.~(\ref{ccons}). The
solution that minimizes the energy must be selected to obtain the ground state.

If one expands the $\cos[(\alpha_{j-1}-\alpha_{j})/2 +\pi(f-n_j)]$ term in 
Eq.~(\ref{ham2}) about it's max\-i\-mum and set $\eta_j=0$, the disc\-rete 
sine-Gordan model (DSG) is obtained:
\begin{equation}
{\cal H} = -J\sum_j \{
	{1\over 2}[(\alpha_{j-1}-\alpha_{j})/2+\pi f]^2 + J_t\cos\alpha_j\},
\end{equation}
A vortex ($n_j=1$) in the JJL corresponds to a kink in the DSG (in the DSG the
$\alpha$ absorb the $n_j$ and are no longer restricted to $(-\pi,\pi]$).  
Kardar \cite{kardar} used this analogy to argue that this system should show 
similar behavior to the DSG which has been studied by 
several authors \cite{aubry,cop,Pokrovsky}. This analogy is only valid for 
$J_t$ very small so that the inter-plaquette term dominates the behavior of the
system making the expansion about its maximum a reasonable assumption.  
However, much of the interesting behavior of the DSG 
occurs in regions of large $J_t$ ($J_t\sim 1$).  Furthermore, much of the work 
by Aubry \cite{aubry} on the DSG relies on the convexity of the 
coupling potential which we do not have in the JJL.

In this paper we formulate the problem in terms of a transfer matrix 
obtained from the full partition function of the ladder.  The eigenvalues 
and eigenfunctions of the transfer matrix are found numerically to
determine the phases of the ladder as functions of $f$ and $J_t$.  We find 
that the spatial periodicity of the ground states goes through a devil's 
staircase as a function of $f$.  We then study the properties of various 
ground states and the low temperature excitations.  As $J_t$ is varied, all 
incommensurate ground states are found to undergo a superconducting-normal 
transition at certain $J_t$ which depends on $f$. In the last section we 
discuss the effects of a current.  

\section{Transfer Matrix Formulation}

The partition function for the ladder, with periodic boundary conditions is
\widetext 
\begin{equation}
{\cal Z} = \prod_i^N \int_{-\pi}^\pi \sum_{\{n_i\}} d\alpha_i d\eta_i 
\exp\left\{K(2 \cos\eta_i \cos[(\alpha_{i-1}-\alpha_i)/2+\pi(f-n_i)]  
+J_t\cos\alpha_i) \right\}.
\label{part}
\end{equation}
where $K=J/k_BT$.  The $\eta_i$ can be integrated out and $n_i$ summed over, resulting in a simple 
transfer matrix formalism for the partition function involving only the 
transverse phase differences:  
${\cal Z} =\prod_i^N \int_{-\pi}^\pi d\alpha_i P(\alpha_{i-1},\alpha_i)
=Tr\, \hat P^N.$
The transfer matrix elements $P(\alpha,\alpha')$ are
\begin{equation}
P(\alpha,\alpha') = 4\pi \exp[KJ_t(\cos\alpha+\cos\alpha')/2] \, 
I_{0}(2 K \cos[(\alpha-\alpha')/2+\pi f]),
\label{mat}
\end{equation}
\narrowtext
where $I_0$ is the zeroth order modified Bessel function 
($I_0(x)={1 \over \pi} \int_0^\pi \exp(x \cos \eta) d\eta$).  Note that
the elements of $\hat P$ are real and positive, so that its largest 
eigenvalue $\lambda_0$ is real, positive and non-degenerate.  However, since
$\hat P$ is not symmetric (except for $f=0$ and $f=1/2$) other eigenvalues 
can form complex conjugate pairs.  As we will see from the correlation 
function, these complex eigenvalues determine the spatial periodicity of the 
ground states. 
  
\begin{figure}
\centerline{\epsfxsize=3in
\epsffile{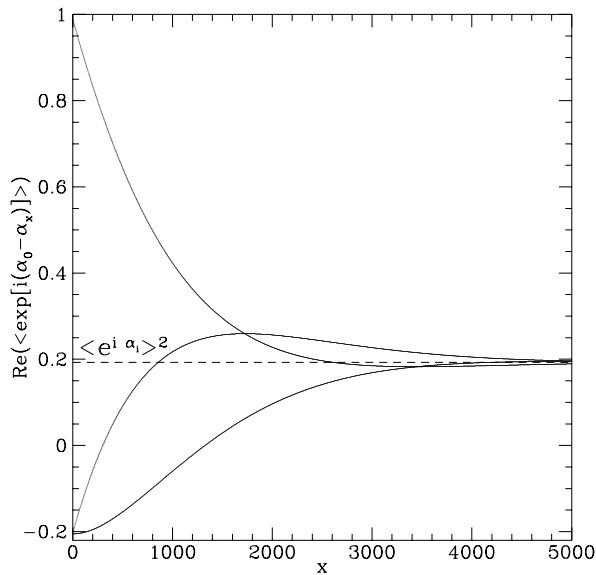}}
\vskip 0.1true cm
\caption{Correlation function for a three periodic state ($\Xi=1/3$).  The 
three branches correspond to $\langle e^{i(\alpha_0-\alpha_{3n})}\rangle$, 
$\langle e^{i(\alpha_0-\alpha_{3n+1})}\rangle$, and 
$\langle e^{i(\alpha_0-\alpha_{3n+2})}\rangle$ with $n=0, 1, 2, \cdots$.  
The dotted line indicates the value of $\langle e^{i \alpha_j}\rangle^2$.}
\label{corfunc}
\end{figure}

The two point correlation function of $\alpha_j$'s is
\begin{eqnarray}
\langle e^{i(\alpha_0-\alpha_l)}\rangle & = & \lim_{N\rightarrow\infty} 
\frac{\left( \prod_{i}^N \int_{-\pi}^\pi d\alpha_{i} 
P(\alpha_{i-1},\alpha_i) \right) e^{i(\alpha_0-\alpha_l)}}{\cal Z}
\nonumber \\
& = & \sum_n c_n \left(\frac{\lambda_n}{\lambda_0}\right)^l,
\label{cor}
\end{eqnarray}
where the $\lambda_n$ are the eigenvalues 
($|\lambda_n|$ $\ge$ $|\lambda_{n+1}|$ and $n=0, 1, 2, \cdots$), 
and the constants 
$c_n=\int_{-\pi}^\pi d\alpha'\psi_0^L(\alpha')
e^{i\alpha'}\psi_n^R(\alpha')\int_{-\pi}^\pi d\alpha\psi_n^L(\alpha)
e^{-i\alpha}\psi_0^R(\alpha)$.  (Note that since $\hat P$ is not sym\-metric 
both right $\psi_n^R$ and left $\psi_n^L$ eigenfunctions are needed.)
If $\lambda_1$ is real and $|\lambda_1|>|\lambda_2|$, Eq.~(\ref{cor}) 
simplifies for large $l$ to
$$
\langle e^{i(\alpha_0-\alpha_l)}\rangle=c_0
                           +c_1\left(\frac{\lambda_1}{\lambda_0}\right)^l,
                           \quad |\lambda_1|>|\lambda_2|.
$$
If $\lambda_1=\lambda_2^*=|\lambda_1|e^{i2\pi\Xi}$, 
Eq.~(\ref{cor}) for large $l$ is
$$
\langle e^{i(\alpha_0-\alpha_l)}\rangle = c_0 + 
  \left(c_1 e^{i2\pi\Xi l}+c_2 e^{-i2\pi\Xi l}\right)\left|\frac{\lambda_1}
 {\lambda_0}\right|^l , \, \lambda_1=\lambda_2^*.
$$
Note that while the correlation length is given by 
$\xi=[\ln|\lambda_0/\lambda_1|]^{-1}$ the quantity $\Xi=Arg(\lambda_1)/2
\pi$ determines the spatial periodicity of the state.  For example, 
Figure~\ref{corfunc} shows the full correlation function (not just the first
two terms) for a three periodic state, $\Xi=1/3$.  We see that the correlation
function has three branches corresponding to 
$\langle e^{i(\alpha_0-\alpha_{3n})}\rangle$, 
$\langle e^{i(\alpha_0-\alpha_{3n+1})}\rangle$, and 
$\langle e^{i(\alpha_0-\alpha_{3n+2})}\rangle$ with $n=0, 1, 2, \cdots$.

It is fairly easy to find the eigenvalues and eigenvectors of the transfer
matrix numerically to very high accuracy (see the Appendix).
For $f$ smaller than a critical value $f_{c1}$ which depends on $J_t$, 
we find that both $\lambda_1$ and $\lambda_2$ are real.  
These two eigenvalues become degenerate at $f_{c1}$, and then bifurcate into a 
complex conjugate pair.  $\Xi$ as a function of $f$ is shown in 
Figure~\ref{stair} for several different values of $J_t$.  The shape of 
these curves is generally referred to as a devil's staircase. 
The steps of the staircase are at $\Xi=p/q$, where $p$ and $q$ are integers. 
These are commensurate states with $p$ vortices in each unit cell which 
consists of $q$ plaquettes.  For small $J_t$, the flat steps are
connected by fairly smooth curves; most states on the $\Xi-f$
curve are incommensurate states.  As $J_t$ increases,  more and more
steps appear and grow at the expense of the smooth regions.  It appears
that at $J_t=J_t^c\approx0.5$ the staircase becomes complete, 
i.e.\ there is a step 
for every rational $\Xi$ and the set of $f$ which correspond to irrational 
$\Xi$ has zero measure.  For $J_t>J_t^c$, the staircase becomes
over-complete, i.e.\ steps of lower order rationals grow and steps of
higher order rationals disappear \cite{foot1}.  A phase diagram can be 
constructed with the phase boundaries at the step edges, as shown in 
Figure~\ref{pbj}.

Another important 
characterization of a state is the phase density $\rho(\alpha)$: 
$\rho(\alpha)d\alpha$ is the average fraction of all sites 
in the ladder with $\alpha<\alpha_i<\alpha+d\alpha$.  If $\rho(\alpha)$
is a smooth function for $\alpha \in (-\pi,\pi]$ at
$T=0$, the ground state energy is invariant under an adiabatic change of 
$\alpha$'s.  Consequently, there is no phase coherence between upper and 
lower branches of the ladder and hence no superconductivity in the
transverse direction.  In this case, we say that the $\alpha$'s are unpinned.  
If there exist finite intervals of $\alpha$ on which $\rho(\alpha)=0$, the 
$\alpha$'s are pinned and there will be phase coherence between the upper and 
lower branches.  In terms of the transfer matrix, the 
phase density is the product of the left and right eigenfunctions of 
$\lambda_0$ \cite{foot2}, $\rho(\alpha)=\psi_0^L(\alpha) \psi_0^R(\alpha)$.

\section{Commensurate States}

We first discuss the case where $f<f_{c1}$.  These are the ``Meissner''
states in the sense that there are no vortices ($n_i=0$) in the ladder.  
The ground state is simply $\alpha_i=0$, $\gamma_j=\pi f$ and $\gamma_j'=-\pi
f$, so that there is a global ``screening'' current $\pm J_x\sin \pi f$ in the
upper and lower branches of the ladder \cite{kardar}.  The phase density
$\rho(\alpha)=\delta (\alpha)$.  
The properties of the Meissner state can be studied by expanding 
Eq.~(\ref{ham2}) around $\alpha_i=0$: 
\begin{equation}
{\cal H}_M = (J/4)\sum_j [\cos(\pi f)(\alpha_{j-1}-\alpha_j)^2
+2 J_t \alpha_i^2].
\label{Hm}
\end{equation}
  At finite temperature, $\rho(\alpha)=\delta (\alpha)$ 
peaks become thermally broadened.   The fluctuations about $\alpha_i=0$ in the 
$J_t\alpha_i^2$ part of the energy of a single plaquette will be of order 
$k_bT$ (from equipartition). Hence $\delta\alpha_i\sim\sqrt{k_bT}$ is an 
estimate of the $\rho(\alpha)$ peak width.
The current conservation Eq.~(\ref{ccons}) becomes
\begin{equation}
\alpha_{j+1}=2 \left( 1+J_t/\cos \pi f\right)\alpha_j-\alpha_{j-1}.
\label{fin_diff}
\end{equation}
Besides the ground state $\alpha_j=0$, there are other two linearly 
independent solutions $\alpha_j=e^{\pm j/\xi_M}$ of Eq.~(\ref{fin_diff}) 
which describe collective fluctuations about the ground state, where
\begin{equation}
{1 \over \xi_M}=\ln\left[1+{J_t \over \cos\pi f}+\sqrt{
    {2 J_t \over \cos\pi f}+\left({J_t \over \cos\pi f}\right)^2}\,\right].
\label{cor-len}
\end{equation}
$\xi_M$ is the low temperature correlation length for the Meissner state.  
(Note that $\xi_M<1$ for $J_t\sim 1$ making a continuum approximation invalid.)
A comparison of $\xi_M$ to the correlation length obtained from the transfer 
matrix (or from an exact solution of the Gaussian model (\ref{Hm}) partition 
function \cite{Parisi}) shows that the two are indistinguishable for 
$k_bT/J \le 0.005$. Note that for finite $J_t$ this model has no zero 
temperature phase transition in the sense that the correlation length remains
finite as $T\rightarrow 0$.  Furthermore, the correlation length diverges 
like $1/\sqrt{2 J_t}$ as $J_t\rightarrow 0$. (or more accurately, approaches 
the rigid rotator value of $1/k_bT$, as we shall see below.)  

As $f$ increases, the Meissner state becomes unstable to the formation of 
vortices.  A vortex is constructed by patching the two vortex free solutions of
 the  Eq.~(\ref{fin_diff}) together anti-symmetrically about the plaquette with
the vortex.  Using Eq.~(\ref{ccons}) for the matching condition,
$$
-\sin (\alpha_*+\pi f)-\sin((e^{1/\xi_M}-1)\alpha_*/2+\pi f)+\sin \alpha_*=0.
$$
where $\pm\alpha_*$ is the the value of $\alpha_j$ on the plaquette enclosing
the vortex. 
Expanded to second order, this gives 
$$
-4 \sin(\pi f) +[2-(e^{1/\xi_M}+1)\cos(\pi f)]\alpha_*+
	[1+(e^{1/\xi_M}-1)^2]\sin(\pi f) \alpha_*^2=0
$$
To first order, the energy $\epsilon_v$ of a single vortex is found to be
$$
\epsilon_v=4 \cos(\pi f)-3 \alpha_* \sin(\pi f)
$$
The zero of $\epsilon_v$ determines $f_{c1}$.  This gives $f_{c1}\approx0.26$
for $J_t=1$.  Extending the calculation of $\epsilon_v$ to second order gives 
$f_{c1}\approx0.28$ for $J_t=1$ which is in good agreement with the numerical 
result from the transfer matrix.  

For $f>f_{c1}$, $\epsilon_v$ is negative and vortices are spontaneously 
created.  When vortices are far apart their interaction is caused only by the 
exponentially small overlap.  The corresponding repulsion energy is of the 
order $J\exp(-l/\xi_M)$, where $l$ is the distance between vortices.  
At finite temperatures and low densities, the vortices will be able to 
overcome a weak pinning potential and will move around fairly freely in the
$l$ sites separating the vortices.
This leads to a free energy per plaquette \cite{Pokrovsky} of 
$$F=\epsilon_v / l + J c \exp(-l/\xi_M)/ l-k_B T \ln l/l,$$
where c is a constant of order unity.  Minimizing this free energy as a 
function of $l$ and putting in the dependence of $\epsilon_v$ near $f_{c1}$ 
gives $\langle n_j \rangle=l^{-1}=[\xi_M\ln|f_{c1}-f|]^{-1}$, for  
$(f-f_{c1})/k_B T >> 1$.

\begin{figure}
\narrowtext
\centerline{\epsfxsize=3in
\epsffile{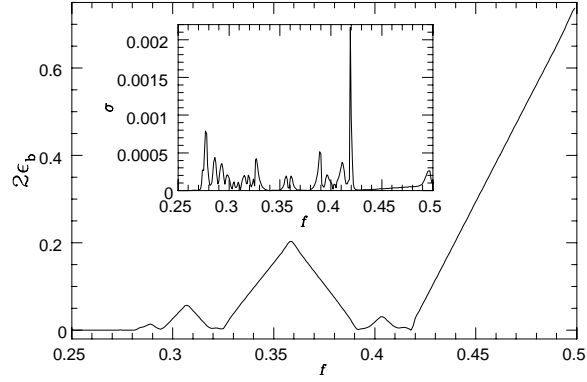}}
\vskip 0.1true cm
\caption{Effective Ising coupling as a function of $f$ for $J_t=1$.  Inset: 
Statistical error for $2\epsilon_b$ in the fit versus $f$.}
\label{tau}
\end{figure}

We now discuss the commensurate vortex states, taking the one with $\Xi=1/2$ as
an example.  This state has many similarities to the Meissner state but some
important differences.  The ground state is 
$$(\alpha_0,n_0)=(\arctan\left[(2/J_t) \sin(\pi f)\right],0),\,\,  
(\alpha_1,n_1)=(-\alpha_0,1);\\
(\alpha_{i\pm 2},n_{i\pm 2})=(\alpha_i,n_i);$$
so that there is a global screening current in the upper and lower branches of 
the ladder of $\pm 2\pi J (f-1/2)/\sqrt{4+J_t^2}$ and the energy is
 $\sqrt{1+(2/J_t)^2 \sin^2(\pi f)}$. 
Global screening, which is absent in an
infinite 2D array, is the key reason for the existence of the steps at
$\Xi=p/q$.  It is easy to see that the symmetry of this $\Xi=1/2$ vortex 
state is that of the (antiferromagnetic) Ising model.  The low temperature 
excitations are domain boundaries between the two degenerate ground states.  
The energy of the domain boundary $J\epsilon_b$ can be estimated using similar 
methods to those used to derive $\epsilon_v$ for the Meissner state.  
We found that $\epsilon_b = \epsilon_b^0 - (\pi^2/\sqrt{4+J_t^2})|f-1/2|$, 
where $\epsilon_b^0$ depends only on $J_t$.
Thus the correlation length diverges with temperature as
$\xi\sim\exp(2J\epsilon_b /k_BT)$.  The transition from the $\Xi=1/2$ state
to nearby vortex states happens when $f$ is such that $\epsilon_b=0$; it is
similar to the transition from the Meissner state to its nearby vortex states. 
All other steps $\Xi=p/q$ can be analyzed similarly.  For comparison, we have
evaluated $\xi$ for various values of $f$ and $T$ from the transfer matrix 
and found that $\xi$ fits $\xi\sim\exp(2 J\epsilon_b/k_BT)$ (typically over 
several decades) at low temperature.  The value of $\epsilon_b$ as a function 
of $f$ is shown in Fig.\ \ref{tau} for $J_t=1$.  The agreement with the above
estimate for the $\Xi=1/2$ step is excellent. 

Another feature that can be fairly easily understood is the relative
heights of the tips of the peaks in Fig.\ \ref{tau} for states with
$\Xi=1/q$.  These states consist of one vortex every $q$ plaquettes in
the ground state configuration.  The lowest energy domain wall
consists of having one spacing that is $q-1$ rather than $q$ between
two consecutive vortices.  This should cost an interaction energy of
about $\tau\sim\exp(-q/l_0)$ with $l_0\approx\xi_M$ of
Eq.~(\ref{cor-len}) and indeed the tips of the peaks in Fig.~\ref{tau}
for states with $\Xi=1/q$ fit this relationship quite well.    

The low temperature 
transverse resistance should be proportional to the thermal activation rate 
of domain walls: $R_t\sim\exp(-2J\epsilon_b /k_BT)$ (with $\epsilon_b$ 
replaced with $\epsilon_v$ for the Meissner state).  This facilitates direct
comparison with experiment \cite{mooij}. 

\section{Incommensurate states and Pinning-Depinning Transition}

We now discuss the incommensurate states and superconducting-normal
transitions in the transverse direction in these states.  An
incommensurate state is a state for which $\Xi$ is an irrational number.
For $J_t=0$, the ground state has $\gamma_i=\gamma_i'=0$ and the
$\alpha_i$ are just pseudo- or slave variables determined by the 
constraint Eq.(\ref{constraint}): 
\begin{equation} 
\alpha_j=2\pi f j+\alpha_0 -2 \pi \sum\nolimits_{i=0}^{i=j} n_i.
\label{j0state}
\end{equation} 
The average vortex density $\langle n_j \rangle$ is $f$; screening currents
are absent.  $\alpha_0$ in Eq.~(\ref{j0state}) is arbitrary; the 
$\alpha$'s are ``unpinned'' for all $f$.  The system is simply two
uncoupled 1D XY chains (or rigid rotor model), so that the correlation
length $\xi=1/k_BT$ (see for instance Ref. \cite{Itzyk}).  The system
is superconducting at zero temperature along the ladder, but not in the
transverse direction.  As $J_t$ rises above zero we observe a distinct 
difference between the system at rational and irrational values of $f$.  For 
$f$ rational, the $\alpha$'s become pinned for $J_t>0$ ($\rho(\alpha)$ is a 
finite sum of delta functions) and the ladder is superconducting in {\it both} 
the longitudinal and transverse directions at zero temperature.  For the $f=0$
case, the correlation length drops from the rigid rotor value of $\xi=1/k_bT$ 
to the value found above in Eq.(\ref{cor-len}) like $T^{-1/2}$. 

\begin{figure}
\narrowtext
\centerline{\epsfxsize=3in
\epsffile{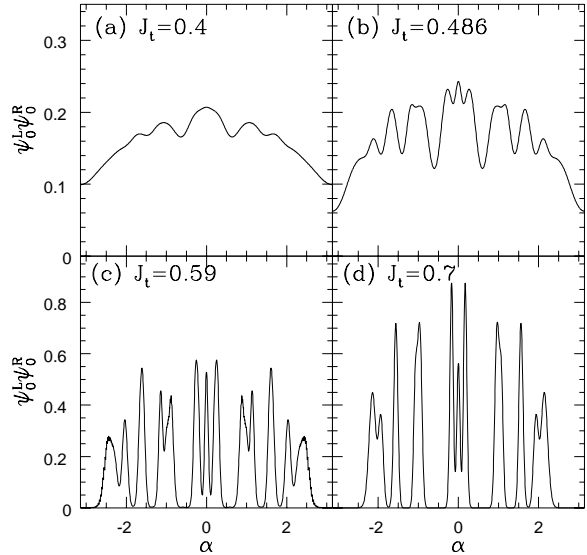}}
\vskip 0.1true cm
\caption{$\rho(\alpha)=\psi_0^L(\alpha)\psi_0^R(\alpha)$ versus $\alpha$ and 
$\Xi$ $=$ $0.381\,966\,011\cdots$, and for (a) $J_t=0.4$, (b) $J_t=0.486$, 
(c) $J_t=0.59$ at $k_BT=0.0007 J$ and (d) $J_t=0.7$ at $k_BT=0.001 J$.  Note 
the smaller scale for the upper plots.}
\label{kam_den}
\end{figure}

\begin{figure}
\narrowtext
\centerline{\epsfxsize=3in
\epsffile{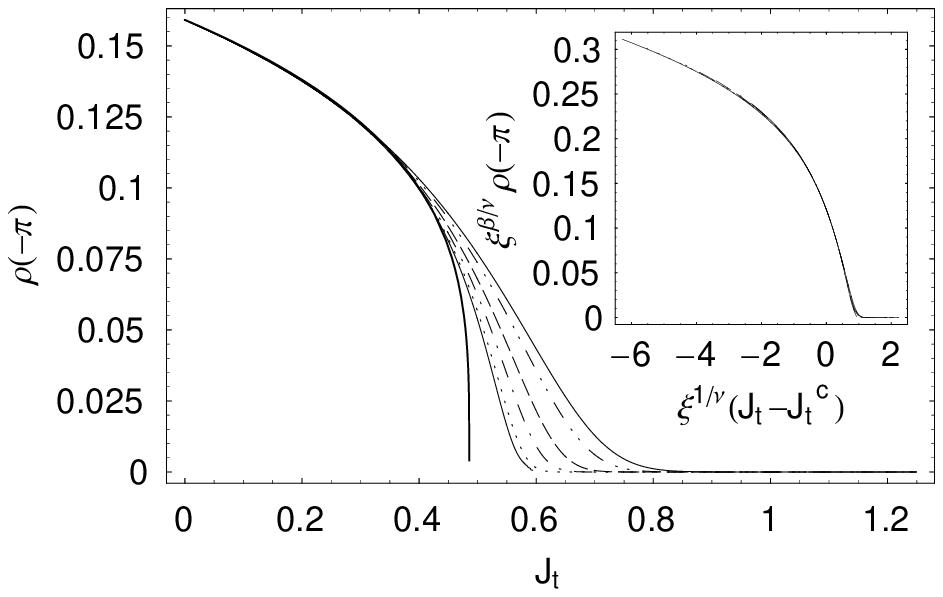}}
\vskip 0.1true cm
\caption{$\rho(-\pi)$ versus $J_t$ for $k_BT=0.012 J$, $0.008 J$, $0.004 J$, 
$0.002 J$, $0.001 J$, and $0.0007 J$.  Also shown is the extrapolation
to $T=0$, $0.192257 (J_t^c-J_t)^\beta$.  Inset:  Scaling collapse of
same data shown in main plot.}
\label{rhopi}
\end{figure}

\begin{figure}
\narrowtext
\centerline{\epsfxsize=3in
\epsffile{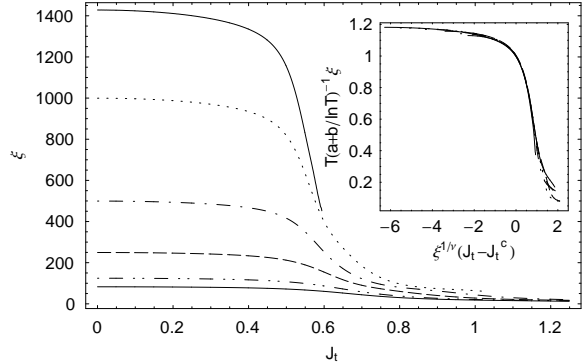}}
\vskip 0.1true cm
\caption{Correlation length $\xi$ versus $J_t$ for $k_BT=0.012 J$, $0.008 J$, 
$0.004 J$, $0.002 J$, $0.001 J$, and $0.0007 J$.  Inset:  Scaling collapse of
same data shown in main plot ($a=0.793\pm 0.006$ and $b=-0.39\pm 0.04$).}
\label{xigm}
\end{figure}

The behavior for irrational $f$ is illustrated in the following for the state 
with $\Xi=a_g$, where $a_g=(3-\sqrt{5})/2$ (one minus the Golden Mean).  
Fig.\ \ref{kam_den} displays $\rho(\alpha)$ for several different $J_t$ at 
$\Xi=a_g$.  We see that the zero-frequency phonon mode (the smoothness of 
$\rho(\alpha)$) persists for small $J_t>0$ until a critical value 
$J_t^c(f)\approx 0.5$ where the $\alpha$'s become pinned and the ladder becomes
 superconducting in the transverse direction.   In the DSG, the pinning 
transition of this state coincides with the devil's staircase of 
Fig.~\ref{stair} becoming complete \cite{aubry,cop} (If the $\alpha_j$'s are 
pinned in this state, then all incommensurate states should be pinned). 
>From Fig.~\ref{kam_den} we see that this transition is a transition where the
integration measure $\rho(\alpha)$ in the partition function goes from a set of
measure one on $[-\pi,\pi)$ to a set of measure zero (at $T\rightarrow 0$).
As one approaches the transition from $J_t<J_t^c$, $\rho(\alpha)\rightarrow 0$
continuously almost everywhere in $[-\pi,\pi)$.  As an order parameter for the
transition, we use $\rho(-\pi)$.  This is shown in Figure~\ref{rhopi} at 
a number of temperatures approaching $T=0$.  
At the same $J_t$ the correlation length drops from the rigid rotor value of 
$\xi=1/(k_BT)$ continuously towards a value of a few tens of lattice 
constants.  This is shown in Figure~\ref{xigm}.  For $J_t<J_t^c$ 
$\xi=1/(k_B T)$ is limited only by the temperature of the system.  This is also
true of the higher order correlation lengths, $[\ln|\lambda_0/\lambda_n]^{-1}$,
 $n>1$ in the unpinned phase which also diverge like $1/(k_B T)$.  

Following standard scaling arguments one would expect $\rho(-\pi)$ to scale 
according to 
$$
\rho(-\pi)(j,\xi)=j^\beta \tilde f(j \xi^{1/\nu}),
$$
where $j=J_t^c-J_t$ and $\xi\sim j^{-\nu}$ for $T=0$ while $\xi$ is a function
of $j$ and $T$ for $T>0$.  For $T\rightarrow 0$, $\xi\rightarrow \infty$ in 
the unpinned phase and $\rho(-\pi)$ should be independent of $\xi$ implying 
$\tilde f(y)\rightarrow constant$ as $y\rightarrow \infty$ so that 
$\rho(-\pi)\sim j^\beta$.  In the opposite extreme, with $T$ fixed and 
$j\rightarrow 0$, $\rho(-\pi)$ is dominated by the finite $\xi$ so that 
$\tilde f(x)\rightarrow 1/x^\beta$ as $x\rightarrow 0$ and 
$\rho(-\pi)\sim \xi^{-\beta/\nu}$.  This allows us to define a more 
convenient scaling function $f(x)=1/x^\beta \tilde f(x)$ so that
$$
\rho(-\pi)(j,\xi)=\xi^{-\beta/\nu} f(j \xi^{1/\nu}).
$$
The results of this scaling, shown in the inset of Figure~\ref{rhopi}, gives
$\beta/\nu=0.0926\pm 0.0009$ and $1/\nu=0.352\pm 0.001$, or
$\beta=0.263\pm 0.003$ and $\nu=2.841\pm 0.008$.

A similar scaling can be applied to $\xi$:
$$
\xi(j,T)=T^{-1}(a+b/\ln T) g(j \xi^{1/\nu})
$$
where we have included the correction to scaling term $b/\ln T$ which causes
a slight improvement to the fit away from $J_t^c$.  (A power law correction
was also considered, but did not have a statistically significant 
coefficient.)  The resulting scaling
collapse is shown in the inset of Fig.~\ref{xigm}.  This does not really
give us any new information, but does act as a check on the information 
from the scaling of $\rho(-\pi)$.

\begin{figure}
\narrowtext
\centerline{\epsfxsize=3in
\epsffile{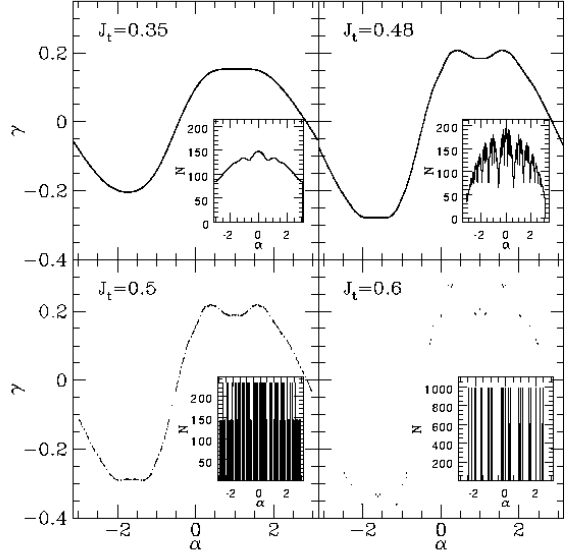}}
\vskip 0.1true cm
\caption{KAM curves constructed from plotting each $(\alpha_j,\gamma_j)$ as
a point in the plane. The state shown is one with periodicity 
$28657/75025\approx0.381\,966\,01$ at $f=28657/75025$. 
Below the critical coupling $J_t^c$, 
the points fill in a smooth curve (top panels) whereas above $J_t^c$ the 
points form a zero measure Cantor set.  Insets: Histograms of the 
$\{ \alpha_j \}$, which can be compared to the phase density $\rho(\alpha)$
calculated from the transfer matrix}
\label{kam}
\end{figure}

The pinning transition of the incommensurate states can be also studied using 
Eq.~(\ref{ccons}) which are equivalent to the two-dimensional map:
\begin{mathletters}
\label{map}
\begin{eqnarray}
\sin \gamma_{j+1} &=& \sin\gamma_j - J_t\sin \alpha_j, \\
\alpha_{j+1} &=& \alpha_j-2\gamma_{j+1}+2\pi(f-n_{j+1}). 
\end{eqnarray}
\end{mathletters}
Vortices are required in order to keep $\alpha_j$ in $(-\pi,\pi]$.  Every 
trajectory of this map is a zero-temperature metastable state of the ladder.
At $J_t=0$ the orbits of (\ref{map}) for the incommensurate states will fill 
in a straight line $\gamma=0$ in the $\gamma\alpha$ plane.  As $J_t$ increases,
 these Kolmogorov-Arnold-Moser (KAM) orbits become deformed but remain 
smooth with the energy remaining independent of $\alpha_0$. Once $J_t$ exceeds 
a critical value the KAM curve breaks down into a Cantor set of measure zero,
as shown in Figure~\ref{kam}.
The disappearance of the zero-frequency phonon mode for irrational $\Xi$'s at 
finite small $J_t^c(f)$ is equivalent to the breakdown of the 
Kolmogorov-Arnold-Moser (KAM) trajectories of the map \cite{KadShen}.  

However,
we have been unable to find any connection between the exponents we found
(describing the approach at $\Xi=a_g$ to $J_t^c$ and $T=0$) to those found
in $\cite{KadShen}$ for the approach to $\Xi=a_g$ at $J_t^c$ and $T=0$ for the
standard map.  If one expands the $\sin \gamma_j\approx \gamma_j$ in 
(~\ref{map}) and redefine $\gamma_j\rightarrow \gamma_j-\pi f$ one obtains the
standard map, studied by several authors including \cite{KadShen,cop}.  
If one compares the critical $J_t^c$ to the equivalent value for the standard
map ($J_t=k_c/2$ in the language of \cite{KadShen}) one finds that they are
identical.  This is somewhat surprising.  One might expect the two problems to
be in the same universality class with related exponents but the location of
the critical point normally depends on the details of the problem.

\section{Effect of a current}

We now turn to the subject of critical currents along the ladder.  One can 
obtain an estimate for the critical current by performing a perturbation 
expansion (i.e. $\{n_j\}$ remain fixed) around the ground state and imposing
the current constraint of 
\begin{equation}
\sin\gamma_j+\sin\gamma_j'=I.
\label{Iconstraint}
\end{equation}
 Letting $\delta\gamma_j$ and $\delta\alpha_j$ be the change of $\gamma_j$ and 
$\alpha_j$ in the current carrying state, an expansion to first order gives
\begin{mathletters}
\label{del}
\begin{eqnarray}
& \cos\gamma_{j-1}\delta\alpha_{j-1}+\cos\gamma_j\delta\alpha_{j+1}\nonumber\\
&  -(\cos\gamma_j+\cos\gamma_{j-1}+2\cos\alpha_j)\delta\alpha_j = 0, 
\label{pur} \\
& \delta\gamma_j = \frac{1}{2}(\delta\alpha_{j+1}-\delta\alpha_j+\frac{I}
{\cos\gamma_j}). \label{dgam}
\end{eqnarray}
\end{mathletters}
Eq.~(\ref{pur}) is in the form of $G\cdot\vec{\delta\alpha}=0$.  If
$\det G \neq 0$, then $\delta\alpha_j=0$ and $\delta\gamma_j=I/2\cos\gamma_j$.
The critical current can be estimated by the requirement that the 
$\gamma_j$ do not pass through $\pi/2$, which gives $I_c = 
2(\pi/2 - \gamma_{\max})\cos\gamma_{\max}$, where $\gamma_{\max}=\max_j
(\gamma_j)$.  In all ground states we examined for $J_t=1.0$, commensurate and
incommensurate, we found that $\gamma_{\max} < \pi/2$, implying a
finite critical current for all $f$.

\begin{figure}
\centerline{\epsfxsize=3in
\epsffile{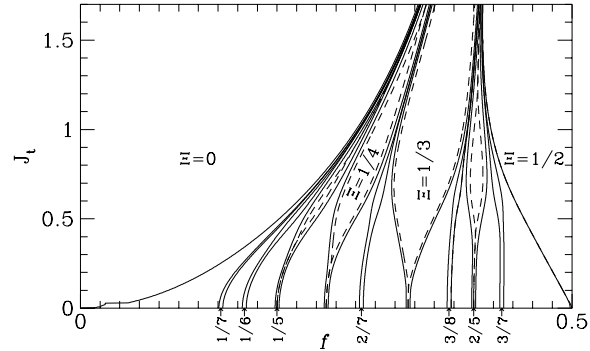}}
\vskip 0.1true cm
\caption{Periodicity $\Xi$ phase diagram for low order rationals.  At 
finite temperature, the step edges are slightly rounded.  The phase
boundaries for a step at $\Xi=p/q$ enclose all $\Xi$ satisfying 
$|\Xi-p/q|<\epsilon$, with dashed lines indicating an $\epsilon$ of $10^{-6}$ 
and solid lines indicating an $\epsilon$ of $0.002$ at $k_BT/J=0.0007$.}
\label{pbj}
\end{figure}

The presence of a current can also have an effect analogous to weakening the
transverse coupling $J_t$.  One can get an approximate phase diagram in the
$I$-$f$ plane, valid for very low temperature and $I<I_c$.
To do this, rather than integrate out the $\eta_i$ in Eq.(\ref{part}), one
can substitute the current constraint Eq.(\ref{Iconstraint}) and get the
``partition'' function for the current carrying states as
\begin{equation}
{\cal Z} = \prod_i^N \int_{-\pi}^\pi  d\alpha_i
2 \exp \left(\beta J_t\cos\alpha_i\right)
\cosh\left[2 K\sqrt{\cos^2[(\alpha_{i-1}-\alpha_i)/2+\pi f]-(I/J)^2/4} \right].
\label{Ipart}
\end{equation}
Again, this is in the form of a product of transfer matrices and can be easily
solved by the same techniques as before (see the appendix).  The resulting 
phase diagram is shown in Figure~\ref{pbi} (compare to Fig.~\ref{pbj}).

Comparing this diagram
to Fig.~\ref{pbj} one can see that the increasing $I$ in Fig.~\ref{pbi}
has the same effect as decreasing $J_t$ in Fig.~\ref{pbj}.  This opens
the possibility of experimentally studying things like the pinning-depinning
transition experimentally. In this case, the pinning-depinning
transition would be driven by the longitudinal current, and a
measurement in the transverse direction would be used to probe the
coherence of the two chains.  

This phase diagram in Fig.~\ref{pbi} should be taken too literally 
however as once the critical current for the system is exceeded, the current
constraint of Eq.(\ref{Iconstraint}) is no longer valid and the system can
switch between metastable states or may change continuously between a very
large number of states.

\section{Conclusion and Acknowledgments}

In conclusion, we have studied the equilibrium behavior of a Josephson junction
ladder in a magnetic field in the absence of charging effects.  Screening
currents play an important role in this system, resulting in the spatial 
periodicity of the ground state climbing a devil's staircase as a function of 
$f$.  Incommensurate states undergo a superconducting-normal transition in 
the transverse direction as $J_t$ is increased, so that for 
$J_t>J_t^c\approx 0.5$ the ladder is superconducting in both the longitudinal 
and transverse directions for all $f$.  The critical current along the 
ladder is found to be finite for all $f$.  Finally, although in one dimension 
there is no phase transition and long range order at finite temperature, 
our study showed that the correlation lengths in vortex states are extremely
long for reasonably low temperatures.  Thus one could test experimentally the
predictions for the vortex configuration by, for instance, direct imaging via a
 scanning Hall-probe microscope or measuring the fractional giant Shapiro steps
\cite{BenzYu}.

We thank Sue Coppersmith, Xinsheng Ling, and Qian Niu for valuable 
discussions.

\appendix
\section{Numerical Solution to Integral Eigenvalue Equations}

\begin{figure}
\centerline{\epsfxsize=3in
\epsffile{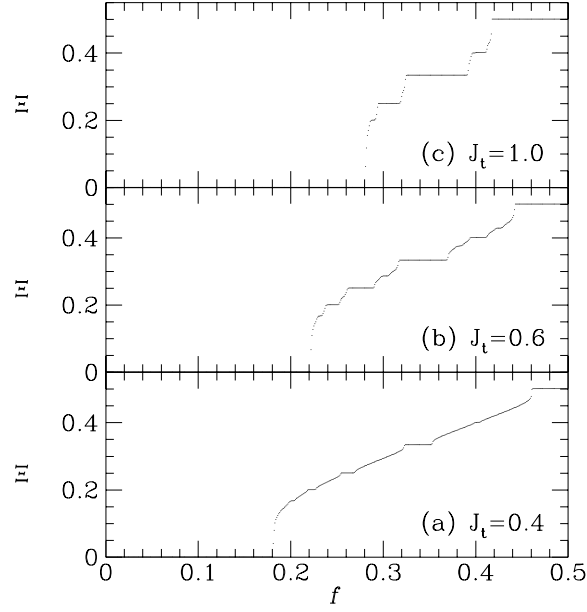}}
\vskip 0.1true cm
\caption{
$\Xi=Arg(\lambda_1)/2\pi$ versus $f$ for $k_BT/J$ $=$ $0.002$ and 
(a) $J_t$ $=$ $0.4$, (b) $J_t$ $=$ $0.6$, and (c) $J_t$ $=$ $1.0$.}
\label{stair}
\end{figure}

\begin{figure}
\narrowtext
\centerline{\epsfxsize=3in
\epsffile{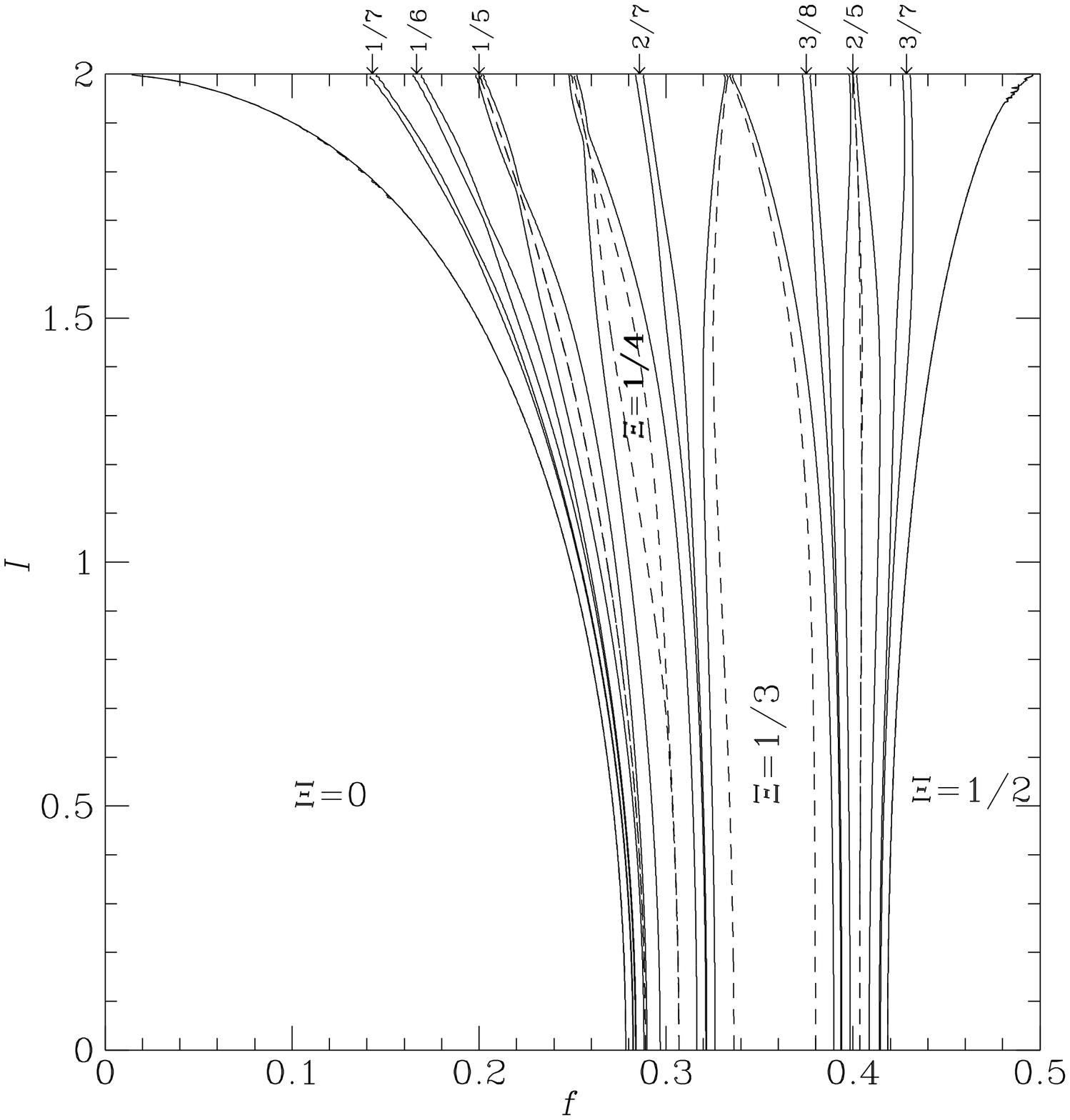}}
\vskip 0.1true cm
\caption{Periodicity as a function of current $I$ and filling factor $f$ for
$J_t=1.0$. At finite temperature, the step edges are slightly rounded.  
The phase boundaries for a step at $\Xi=p/q$ enclose all $\Xi$ satisfying 
$|\Xi-p/q|<\epsilon$, with dashed lines indicating an $\epsilon$ of $10^{-6}$ 
and solid lines indicating an $\epsilon$ of $0.002$ at $k_BT/J=0.004$.
Note that some of the ``tongues'', such as the one for $\Xi=1/4$ come
to a constriction at what appears to be the critical current for that
state.  The section above the constriction is, therefore, above the
critical current and should not be taken too seriously.}
\label{pbi}
\end{figure}

To find the eigenvalues of the transfer matrix (\ref{mat}), we used the Nystrom
 method with n-point Gauss-Legendre quadrature \cite{DelvesMohamed}. 
 The resulting matrix eigenvalue problem was then solved using the 
routines in LAPACK.  It is worth describing this method here as it is fairly 
simple.  We should also note that this method is quite fast, compared to 
methods such as the effective potential methods used by Mazo et al. in 
Ref.~\cite{laddergroups}.  Indeed one can generate phase diagrams such
as those in Fig.~\ref{stair} in a few minutes or in Fig.~\ref{pbi} in a
hour or so of computer time (on an SGI workstation).

 The transfer matrix $\hat P$ is not symmetric (except when $f=0$ or $1/2$), 
and we will therefore require both right $\psi_n$ and left $\tilde \psi_n$ 
eigenvectors for the calculation of correlation functions.  The eigenvalues
and left and right eigenvectors are defined by the relations
\begin{eqnarray}
\int_{-\pi}^\pi d\alpha' P(\alpha,\alpha')\psi_n(\alpha') &=&
\lambda_n \psi_n(\alpha) \qquad (|\lambda_n| \geq |\lambda_{n+1}|) \nonumber\\
\int_{-\pi}^{\pi} d\alpha \tilde \psi_n(\alpha) P(\alpha,\alpha')
&=& \lambda_n \tilde \psi_n(\alpha'). 
\label{ev_prob}
\end{eqnarray}
The left and right eigenvectors are orthonormal ($\int d\alpha \,\tilde\psi_n
(\alpha) \psi_{n'}(\alpha)=\delta_{nn'}$) and complete (In all cases examined,
 the eigenvalues, though possibly equal in magnitude as in a complex conjugate 
pair, were nondegenerate).  Note that the elements of the transfer matrix 
$\hat P$ are real and positive, so that the largest eigenvalue is real, 
positive and nondegenerate.  Also, any complex eigenvalues will occur in 
complex conjugate pairs so that the partition function remains real and 
positive for any $N$.

The Nystrom method requires the choice of some approximate quadrature rule:
\begin{equation}
\lambda_n \psi_n(\alpha)= \int_{-\pi}^{\pi} d\alpha' P(\alpha,\alpha')\psi_n(\alpha') \approx 
\sum_{j=1}^n w_j P(\alpha,\alpha'_j) \psi_n{\alpha'_j}
\label{quadrule}
\end{equation}
Here the set $\{w_j\}$ are the weights of the quadrature rule, while the $n$
points $\{\alpha'_j\}$ are the abscissas.  Since the solution method involves
$O(n^3)$ operations it is worthwhile to use a high-order quadrature rule.  
For smooth, nonsingular problems Gaussian quadrature tends to be the best.
Elements of the transfer matrix become increasingly singular as 
$T\rightarrow 0$, which produces computational difficulties. (For instance, a 
free energy per plaquette of 2.836 at $k_bT/J=0.004$ corresponds to
$\lambda_0\sim 10^{308}$, or close to overflow on the Silicon Graphics Indigo 
workstation used).  Despite this problem, we found n-point Gauss-Legendre
quadrature (n was typically 150 to 500) to give quite reasonable results.  The 
largest errors tend to occur in the location of the edge of the steps of the 
devil's staircase, and even these errors are far too small to be visible on the
 plots. The lowest temperature achieved was $k_bT/J=0.0007$, which we
 believe is sufficient to characterize the states of the zero temperature 
system (to achieve this temperature, a constant was added to the energy so 
that the ground state energy was near zero, thus helping to correct for the
overflow problem). 

Equation (\ref{quadrule}) is a standard eigenvalue equation
\begin{equation}
\hat{\bf P} \cdot {\bf \psi}=\lambda {\bf \psi}
\end{equation}
which was solved using the routines in LAPACK (LAPACK library routines are
provided on most workstation based systems and source code can be obtained
free from netlib \cite{netlib}).  When making use of the eigenvectors (such
as in the calculation of correlation functions) at points not included in the 
quadrature points one should make use of Eq.(\ref{quadrule}) in order
to maintain the full accuracy of the solution.  In addition, in order to keep
track of the weights, it is useful to symmetrize the weighting: defining the 
diagonal matrix ${\bf D}^{1/2}=diag(\sqrt{w_j})$. then the eigenvalue equation 
becomes
$$
( {\bf D}^{1/2} \hat{\bf P} {\bf D}^{1/2}) \cdot ({\bf D}^{1/2}{\bf \psi})=
\lambda ({\bf D}^{1/2}{\bf \psi}),
$$
which is in the form of a symmetric eigenvalue problem for $f=0,1/2$.

\begin{references}
\bibitem[*]{present}
Present Address: Dept. of Physics, Theoretical Physics, University of
Oxford, 1 Keble Road, Oxford OX1 3NP.
\bibitem{jja}
For a general review, see Physica B {\bf 152}, 1-302 (1988).
\bibitem{us1}
C. Denniston and C. Tang, Phys. Rev. Lett. {\bf 75}, 3930 (1995).
\bibitem{kardar}
M. Kardar, Phys. Rev. B {\bf 30}, 6368 (1984); Phys. Rev. B {\bf 33}, 3125 
(1986).
\bibitem{laddergroups}
E. Granato, Phys. Rev. B {\bf 42}, 4797 (1990); 
S. Ryu, W. Yu, and D. Stroud, Phys. Rev. E {\bf 53}, 2190, (1996);  
J. Mazo, F. Falo and L. Floria, Phys. Rev. B {\bf 52}, 10\,433 (1995);
J. Mazo and J. Ciria, Phys. Rev. B {\bf 54}, 16068 (1996).
\bibitem{mooij}
H. Eikmans {\it et al.}, Physica B {\bf 165}\&{\bf 166}, 1569 (1990); 
H. Eikmans and J. van Himbergen, Phys. Rev. B {\bf 41}, 8927 (1990).
\bibitem{anderson}
P. W. Anderson, in {\it Lectures on the Many Body Problem}, edited by E.R. 
Caianiello (Academic Press, New York, 1964), Vol. 2, pp. 113.
\bibitem{tinkham}
M. Tinkham, {\it Introduction to Superconductivity}, (Mc\-Graw-Hill, New 
York, 1975).
\bibitem{aubry}
S. Aubry and G. Andr\'e, in {\it Group Theoretical Methods in Physics}, 
edited by Horowitz and Ne'eman, Ann. Israel Phys. Soc. {\bf 3}, 133 (1980); 
S. Aubry, Ferroelectrics {\bf 24}, 53 (1980); 
in {\it The Riemann Problem, Complete Integrability and 
Arithmetic Applications}, Vol. 925 of {\it Lecture Notes in Mathematics} 
(Springer-Verlag, 1982), pp. 221-245.
\bibitem{cop}
S.N. Coppersmith and D.S. Fisher, Phys. Rev. B {\bf 28}, 2566 (1983).
\bibitem{Pokrovsky}
V.L. Pokrovsky, A.L. Talapov and P. Bak, in {\it Solitons}, edited by 
Trullinger {\it et al.}, (Elsevier, 1986), pp. 71-127.
\bibitem{foot1}
We are grateful to S. Coppersmith for pointing out to us that in the limit of 
$J_t \rightarrow \infty$, only integer steps survive, so that $\Xi=0$ for 
$f<1/2$ and $\Xi=1$ for $1/2<f<1$.
\bibitem{foot2}
R.A. Guyer and M.D. Miller, Phys. Rev. Lett. {\bf 42}, 718 (1979).
\bibitem{Parisi}
see for instance, G. Parisi, {\it Statistical Field Theory}, 
(Addison-Wesley,1988), pp. 227.
\bibitem{Itzyk}
see for instance, Itzykson and Drouffe, {\it Statistical Field Theory} 
(Cambridge Univ. Press, 1989), pp. 40.
\bibitem{KadShen}
For scaling analysis of the breakdown of KAM trajectories in the
``standard map'', see L.P. Kadanoff, Phys. Rev. Lett. {\bf 47}, 1641 (1981) 
and S.J. Shenker and L.P. Kadanoff, J. Stat. Phys. {\bf 27}, 631 (1982).
\bibitem{BenzYu}
S.P. Benz {\it et al.}, Phys. Rev. Lett. {\bf 64}, 693 (1990); 
W. Yu {\it et al.}, Phys. Rev. B {\bf 45}, 12624 (1992).
\bibitem{DelvesMohamed}
L.M. Delves and J.L. Mohamed, {\it Computational Methods for Integral 
Equations} (Cambridge Univ. Press, 1985);  or W.H. Press et. al., 
{\it Numerical Recipes, 2nd Edition} (Cambridge Univ. Press, 1992).
\bibitem{netlib}
see http://csep1.phy.ornl.gov/cornell\_proceedings/tutorials/Xnetlib/xnetlib\_ov.html .
\end{references}
\end{document}